\documentclass[twocolumn,prl,aps,longbibliography,10pt]{revtex4-1}

\usepackage{amsfonts,amssymb}
\usepackage[sumlimits,intlimits]{amsmath}
\usepackage{graphics}
\usepackage{graphicx}
\usepackage{mathrsfs}
\usepackage{textcomp}
\usepackage{verbatim}
\usepackage{hyperref}    
\usepackage{bm}          

\newcommand{\be}{\begin{equation}}
\newcommand{\ee}{\end{equation}}
\newcommand{\e}{\varepsilon}
\newcommand{\rr}{\mathbf{r}}
\newcommand{\kk}{\mathbf{k}}
\newcommand{\PP}{\mathbf{P}}
\newcommand{\he}{\widehat{\e}}
\newcommand{\qq}{\mathbf{q}}

\newcommand{\pt}{\widetilde{\psi}}
\newcommand{\Vt}{\widetilde{V}}
\newcommand{\phit}{\widetilde{\phi}}

\begin{document}

\title{Interlayer excitons with tunable dispersion relation}

\author{Brian Skinner}
\affiliation{Massachusetts Institute of Technology, Cambridge, MA 02139  USA}

\date{\today}

\begin{abstract}

Interlayer excitons, comprising an electron in one material bound by Coulomb attraction to a hole in an adjacent material, are composite bosons that can assume a variety of many-body phases.  The phase diagram of the bosonic system is largely determined by the dispersion relation of the bosons, which itself arises as a combination of the dispersion relations of the electron and hole separately.  Here I show that in situations where either the electron or the hole has a non-monotonic, ``Mexican hat-shaped" dispersion relation, the exciton dispersion relation can have a range of qualitatively different forms, each corresponding to a different many-body phase at low temperature.  This diversity suggests a novel platform for continuously tuning between different quantum phases using an external field.

\end{abstract}

\maketitle

When an electron binds to a hole in a solid state system, the resulting exciton has properties that are qualitatively different from either the electron or hole separately.  For example, the electron and hole have fermionic statistics, while the exciton is a boson.  Electrons and holes interact via a long-ranged Coulomb interaction, while excitons have only a short-ranged dipolar interaction.  But in terms of its dispersion relation, an exciton is usually qualitatively similar to a free electron or hole.  In particular, one can usually describe the center of mass coordinate of the exciton as an effectively free particle having mass equal to the sum of the electron and hole masses, so that the total energy $\e$ of the exciton grows as $P^2$, where $P$ is the center of mass momentum.

But what happens when the electron and hole have dispersion relations that are qualitatively different from each other?  Which of its constituent particles does the exciton take after, in terms of its dispersion relation: the electron or the hole?

\begin{figure}[htb]
\centering
\includegraphics[width=0.48 \textwidth]{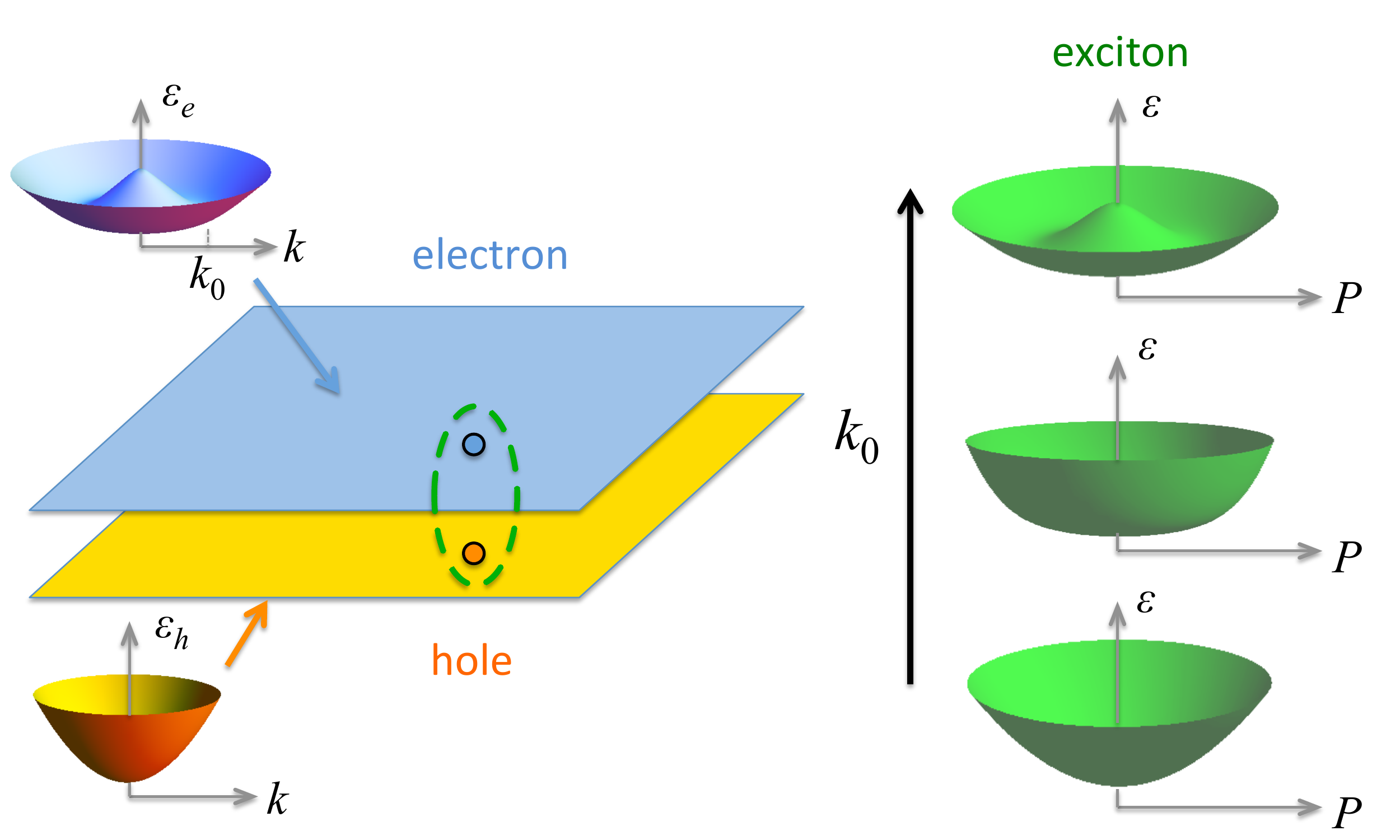}
\caption{(Color online) Schematic illustration of a composite boson with tunable dispersion relation. A material containing electrons (blue layer) is placed adjacent to a material containing holes (yellow layer), with the electron dispersion relation following Eq.\ (\ref{eq:ee}) and the hole dispersion relation following Eq.\ (\ref{eq:eh}).  In this configuration the electron and hole can bind together to form an interlayer exciton, which has a dispersion relation that can take a number of qualitatively different forms, depending on the value of $k_0$.}
\label{fig:schematic}
\end{figure}

In particular, consider the case of a two-dimensional (2D) exciton for which the electron has a ``Mexican hat-shaped" dispersion, 
\be 
\e_e \left(\kk \right) = \frac{\left(\left|\kk \right| - k_0 \right)^2}{2m_e},
\label{eq:ee}
\ee 
while the hole has the usual parabolic dispersion
\be 
\e_h\left(\kk \right) = \frac{\left|\kk \right|^2}{2m_h}.
\label{eq:eh}
\ee 
Here, $m_e$ and $m_h$ are the effective masses of the electron and hole, respectively, $\kk$ is the momentum, and $k_0$ is a characteristic momentum that defines the width of the ``brim" of the Mexican hat (see Fig.\ \ref{fig:schematic}).  
One can now ask the question: what is the form adopted by the exciton dispersion relation $\e(P)$?  

This seemingly quaint problem actually has a range of experimental implications.  As explained below, its solution suggests the ability to engineer interlayer excitons with widely tunable dispersion relation.\footnote{This tunability is similar in spirit to the problem of mobile magnetic impurities in a superfluid\cite{gopalakrishnan_mobile_2015}}  This tunability offers the potential to realize a range of many-body quantum phases within a single device by changing an external field.

The idea of using interlayer excitons to realize a 2D Bose system goes back more than forty years,\cite{lozovik_feasibility_1975} but its experimental realization has been enabled only relatively recently by the development of sufficiently clean and sufficiently thin bilayer devices.\cite{butov_anomalous_1998, eisenstein_boseeinstein_2004, butov_macroscopically_2002}  The ongoing development of new 2D materials continues to provide novel platforms for realizing bilayer exciton physics.\cite{zhang_excitonic_2008, seradjeh_exciton_2009,  rivera_observation_2015}
Of particular significance is the identification of a number of different materials that have, or can be made to have, a Mexican hat shape at low energy $\e$.  Such a dispersion relation is usually associated with Rashba spin-orbit coupling,\cite{manchon_new_2015} which can be significant in semiconductors like GaAs or InAs,\cite{wang_electrical_2015, nitta_gate_1997, grundler_large_2000} at the surface of topological insulators like Bi$_2$Se$_3$,\cite{king_large_2011} or at oxide interfaces such as LaAlO$_3$/SrTiO$_3$.\cite{caviglia_tunable_2010} A Mexican hat-shaped dispersion also arises in materials such as bilayer graphene\cite{mccann_landau-level_2006, mccann_electronic_2013} that have an avoided crossing between two intersecting bands with opposite-sign velocity.  Importantly, for each of the preceding examples in this paragraph, the brim of the Mexican hat, $k_0$, can be widely and continuously adjusted by applying a transverse electric field.  As shown below, this adjustability offers the ability to continuously tune the dispersion relation of interlayer excitons.  The remainder of this paper is dedicated to deriving the exciton dispersion relation, and to discussing the different many-body phases that can be obtained by tuning $k_0$ and the exciton density $n$.  For simplicity, the amplitude of inter-layer tunneling is considered everywhere to be negligibly small.

Before giving a detailed calculation of the exciton dispersion, it is worth outlining how the dispersion relation can be understood qualitatively using the following simple scaling arguments, which for simplicity assume small interlayer separation $d$.  The key idea is to compare $k_0$ with the typical internal momentum $k_e$ of the electron within a stationary exciton.  When $k_0$ is much smaller than $k_e$, the (small) Mexican hat feature in the electron dispersion is irrelevant, since the electron wave function uses primarily much larger momentum components.  When $k_0 \gg k_e$, on the other hand, the Mexican hat structure in $\e_e(k_e)$ is reflected in the exciton dispersion relation, since the energy of the electron is strongly reduced when the exciton acquires enough momentum that the electron momentum approaches $k_0$.

Consider first the case where the hole mass is heavy, $m_h \gg m_e$.  In this case the characteristic radius $r_h$ of the hole wavefunction is much smaller than that of the electron, $r_e$.  Consequently, $r_e$ is similar to the radius of a 2D hydrogen atom with a stationary nucleus, $r_e \approx \kappa/m_e e^2 \equiv a_e$ (the electron effective Bohr radius, in Gaussian units with $\hbar \equiv 1$ --- here $\kappa$ is the dielectric constant).  This expression for $r_e$ is roughly correct even when $k_0$ is large, since large $k_0$ leads only to a logarithmic renormalization of $r_e$.\cite{skinner_bound_2014}  The typical electron momentum is therefore $k_e \sim 1/a_e$.  If $k_0$ is small enough that $k_0 \ll 1/a_e$, then the non-monotonic part of the electron dispersion at small $\kk$ is smeared out by the internal motion of the electron around the hole, and the exciton dispersion is parabolic.  Thus, it is only at larger $k_0$, such that
\be 
k_0 a_e \gg 1, \hspace{5mm} (m_h/m_e \gg 1),
\label{eq:boundarylargemh}
\ee 
that the exciton dispersion has a Mexican-hat shape. 

On the other hand, when the hole mass is light enough that $m_h \ll m_e$, the hole wavefunction acquires a large size $r_h \approx \kappa/m_h e^2$, while the electron wavefunction is relatively compact.  The spatial extent of the electron wavefunction in this case can be found by considering that the electron sits in the bottom of a parabolic potential well created by the Coulomb potential of the hole.  The shape of this potential well is $u(\rho) \sim e^2 \rho^2/\kappa r_h^3$, where $\rho$ is the displacement of the electron from the center of the potential well, and the corresponding size of the electron wavefunction is that of the ground state of the 2D harmonic oscillator --- namely, $r_e \sim (a_e r_h^3)^{1/4}$.  (As in the previous case, this expression for $r_e$ is correct even for large $k_0$.)  The Mexican hat shape of the dispersion relation remains  only when $k_0 \gg 1/r_e$, which is equivalent to the condition
\be 
k_0 a_e \gg \left(\frac{m_h}{m_e} \right)^{3/4}, \hspace{5mm} (m_h/m_e \ll 1).
\label{eq:boundarysmallmh}
\ee 

These qualitative predictions can be verified in a quantitative way as follows.  The exciton dispersion $\e(P)$ is found by solving the Schrodinger equation $H \psi = \e \psi$ with the Hamiltonian
\be 
H = \he_e + \he_h - V(\rr_{eh}),
\ee 
where $\he_e$ and $\he_h$ are the kinetic energy operators for the electron and hole, respectively, and
\be 
V(\rr_{eh}) = - \frac{e^2}{\kappa \sqrt{|\rr_{eh}|^2 + d^2}}
\ee 
is their mutual Coulomb energy.  Here $\rr_{eh} = \rr_e - \rr_h$ is the displacement vector between the electron and hole, with $\rr_e$ being the electron coordinate and $\rr_h$ the hole coordinate.  The wavefunction with fixed total momentum $\PP$ is written as
\be 
\psi(\rr_e, \rr_h) = \exp[i (\kk_e^0 \cdot \rr_e + \kk_h^0 \cdot \rr_h)] \varphi(\rr_{eh}),
\label{eq:varpsi}
\ee 
where $\kk_e^0$ and $\kk_h^0$ are wave vectors such that $\kk_e^0 + \kk_h^0 = \PP$ and $(d\e_e/d\kk)|_{\kk=\kk_e^0} = (d\e_h/d\kk)|_{\kk=\kk_h^0}$.  These two conditions ensure that the electron and hole together have total momentum $\PP$, and that they both have the same group velocity.  (Equivalently, $\kk_e^0$ and $\kk_h^0$ are the values of momentum that minimize the total kinetic energy $\e_e(\kk_0^e) + \e_h(\kk_0^h)$ under the constraint $\kk_e^0 + \kk_h^0 = \PP$.)  One can say that the distribution of electron momenta is centered around $\kk_e^0$, while the hole has momenta in the neighborhood of $\kk_h^0$.  For concreteness, below I take $\PP$ to be in the $x$ direction, so that $\kk_e^0$ and $\kk_h^0$ are also in the $x$ direction.

The function $\varphi(\rr_{eh})$ represents the wave function for the relative motion of the electron and hole around each other.  Such motion is, in general, anisotropic, corresponding to a hydrogen-like state that is elongated in one direction.  One can see how this anisotropy arises by considering the special case in which $k_0$ is large and $m_e$ is small.  In this case the electron kinetic energy has a sharp minimum at $|\kk_e| = k_0$, and $\kk_e^0$ is very close to $k_0 \hat{x}$.  The dispersion relation for the electron in the neighborhood of $\kk_e^0$ is therefore parabolic in the $x$ direction with mass $m_e$, and nearly flat in the $y$ direction.  Consequently, the electron wavefunction acquires a shape that is tightly confined in the $y$ direction (extended in momentum space) and more extended in the $x$ direction (compact in momentum space).

In order to account for this anisotropy, one can use the variational wavefunction
\be 
\varphi(x, y) = \sqrt{\frac{2 \beta^2 \lambda}{\pi}} \exp[-\beta \sqrt{x^2 + (\lambda y)^2} ]
\label{eq:psirel}
\ee 
(following, for example, Refs.\ \onlinecite{kittel_theory_1954, prada_effective-mass_2015}).
Here, $\beta$ and $\lambda$ are variational parameters, with $\beta$ corresponding to the inverse size of the wavefunction in the $x$ direction, and $\lambda > 1$ being the dimensionless anisotropy of the wave function.  In the limiting case $k_0 \rightarrow 0$ and $d \rightarrow 0$, Eq.\ (\ref{eq:psirel}) reproduces the ground state wavefunction of the hydrogen atom at $\lambda = 1$.  More generally, the energy of the exciton is approximated by the minimum value of $\langle \psi | H | \psi \rangle$ over all values of the variational parameters.  That is,
\be 
\e(P) \simeq \min_{ \beta, \lambda} (\e_k + \e_c),
\ee 
where $\e_k$ is the expectation value of the kinetic energy, given by
\be 
\e_k = \frac{32 \beta^4}{\pi \lambda} \int d^2\qq \frac{ \e_e(\kk_e^0 + \qq/2) + \e_h(\kk_h^0 - \qq/2)}{\left[ 4 \beta^2 + q_x^2 + q_y^2/\lambda^2 \right]},
\label{eq:ek}
\ee 
and $\e_c$ is the Coulomb energy of the exciton, given by
\be 
\e_c \simeq \frac{-2 e^2 \beta^2 \lambda}{\pi \kappa} \int d^2\rr \frac{\exp[-2 \beta \sqrt{x^2  + (\lambda y)^2}]}{\sqrt{x^2 + y^2 + d^2}}.
\label{eq:ec}
\ee 
Equations (\ref{eq:ek}) and (\ref{eq:ec}) are derived in the Supplementary Information, along with an analytical expression for Eq.\ (\ref{eq:ec}) in the limit $d/a_e \rightarrow 0$.

\begin{figure}[htb]
\centering
\includegraphics[width=0.5 \textwidth]{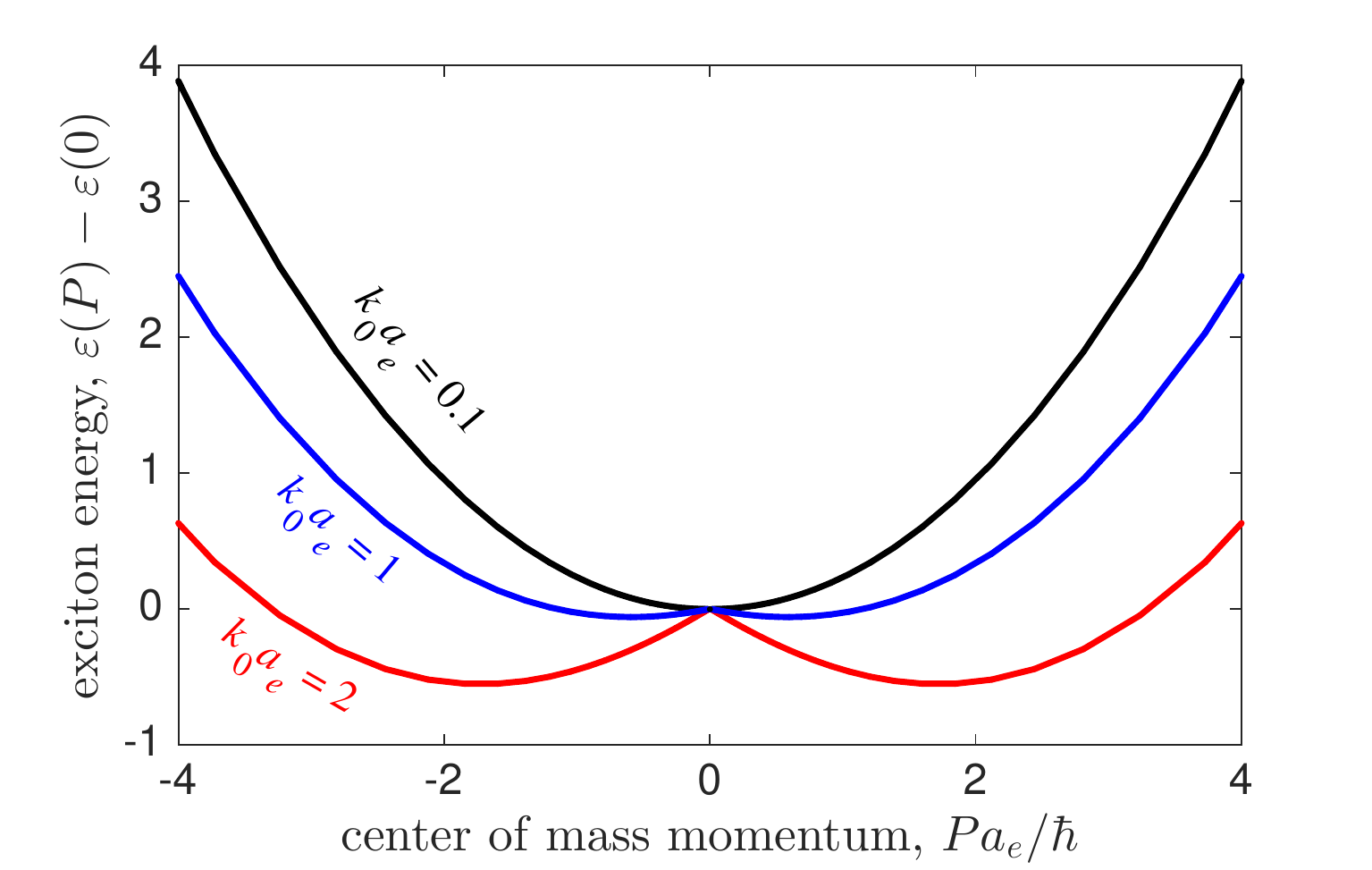}
\caption{(Color online) The dispersion relation $\e(P)$ of an interlayer exciton with $m_h = m_e = 1$ and $d/a_e \rightarrow 0$.  Different curves are labeled by the corresponding value of $k_0$. Units of energy on the vertical axis are $e^2/\kappa a_e$.}
\label{fig:dispersions}
\end{figure}

Figure \ref{fig:dispersions} shows an example calculation of the exciton dispersion relation $\e(P)$ for different values of $k_0$, taking the case of $m_h = m_e$ and $d/a_e \rightarrow 0$.  For small $k_0 \ll 1/a_e$, the dispersion is parabolic, and well described by $\e(P) = P^2/[2(m_e + m_h)]$.  At large $k_0 \gg 1/a_e$, on the other hand, the exciton has a pronounced minimum in the dispersion at $|\PP| \simeq k_0$, and $\e(P) \simeq (|\PP| - k_0)^2/[2(m_e + m_h)]$.  In between these two extremes, at $k_0a_e \approx 1$, the exciton dispersion relation becomes relatively flat at small momentum, resembling $\e(P) \simeq P^4/(2 m k_0^2)$ for momenta $|\PP| \lesssim k_0$.  

\begin{figure}[htb]
\centering
\includegraphics[width=0.48 \textwidth]{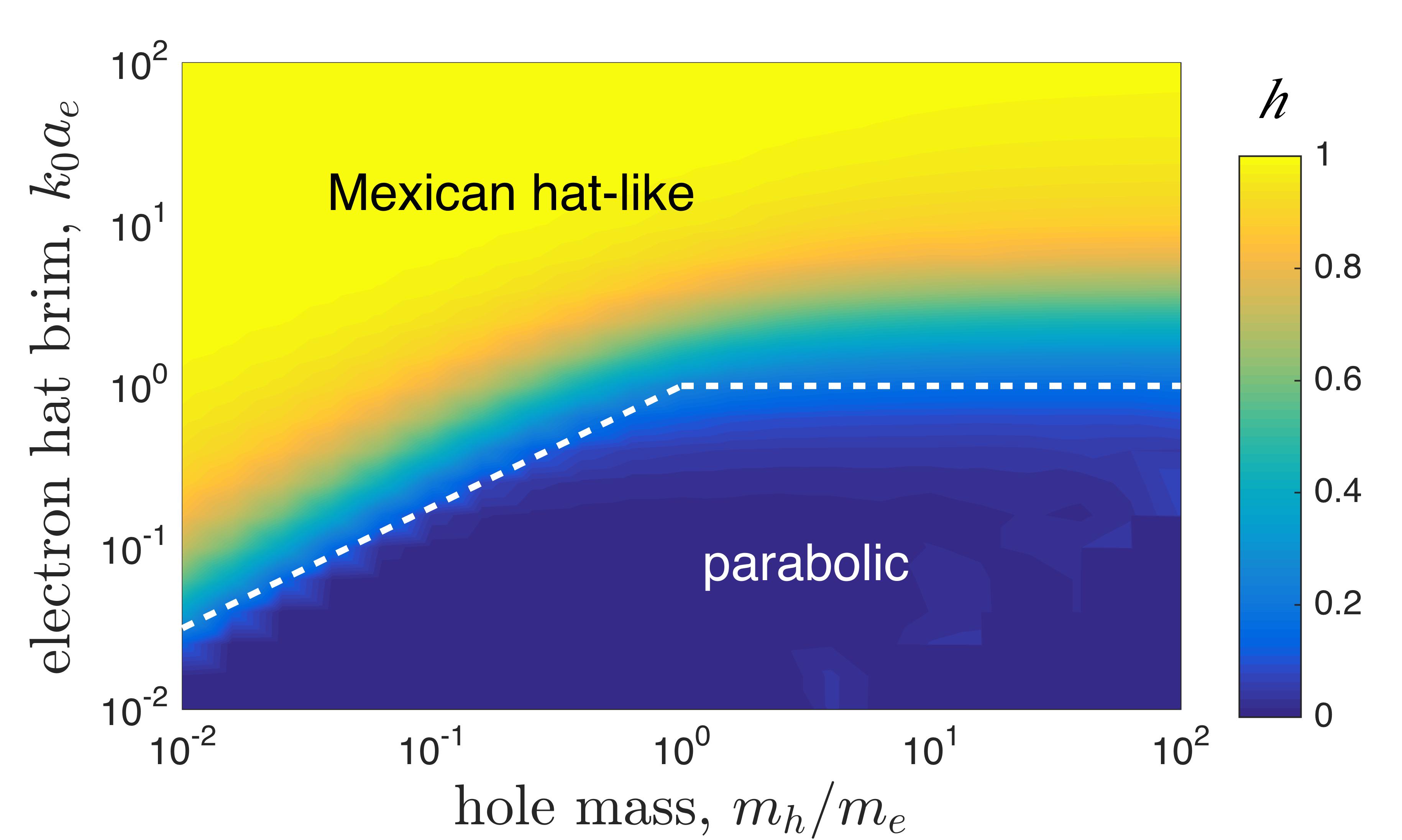}
\caption{(Color online) The normalized hat height of the exciton dispersion, $h = [\e(P=0)-\e_\text{min}]/[k_0^2/(2(m_e+m_h))]$, as a function of the electron ``hat brim" $k_0$ and the mass ratio $m_h/m_e$.  Regions of $h \approx 1$ are such that the exciton dispersion relation is Mexican hat-like, $\e(P) \simeq (P - k_0)^2/(2(m_e + m_h))$, while regions of $h \approx 0$ are such that $\e(P) \simeq P^2/(2(m_e + m_h))$.  The dashed white lines indicate the approximate boundary between these two regions derived in Eqs.\ (\ref{eq:boundarylargemh}) and (\ref{eq:boundarysmallmh}).}
\label{fig:hatheight}
\end{figure}

One can explore the evolution of the dispersion relation in a more systematic way by evaluating $\e(P)$ for a range of values of $k_0$ and $m_h/m_e$.  The degree of Mexican hat-ness for each case can be quantified by making the ratio $h = [\e(P=0) - \e_\text{min}]/[k_0^2/(2(m_e + m_h))]$, where $\e_\text{min}$ is the minimum energy of the exciton as a function of momentum $P$.  $h$ can be called the ``normalized hat height" of the exciton, with $h = 0$ indicating a parabolic dispersion relation and $h = 1$ indicating a Mexican hat-shaped dispersion.  The normalized hat height is plotted in Fig.\ \ref{fig:hatheight} for a wide range of the parameters $k_0 a_e$ and $m_h/m_e$.  Its behavior is in close agreement with the qualitative derivation presented in Eqs.\ (\ref{eq:boundarylargemh}) and (\ref{eq:boundarysmallmh}).

One can now consider the implications of this tunable dispersion for a large system of excitons with finite number density $n$.  For such a system, altering the dispersion relation can imply a change to the ground state phase.  Consider, for example, that when $k_0 = 0$, the system is equivalent to a 2D collection of dipolar bosons with finite mass $m = m_e + m_h$.  Such a system can occupy either a liquid phase (a Bose-Einsten condensate or superfluid) or a solid phase (a Wigner crystal), depending on the value of the density.\cite{babadi_universal_2013, astrakharchik_quantum_2007}  For point-dipoles, the solid phase exists at $n \gg \kappa^2/(m^2 e^4 d^4)$. (This condition comes from the requirement that the typical nearest-neighbor interaction energy $\sim e^2 d^2 n^{3/2}/\kappa$ is larger than the corresponding quantum confinement energy $\sim n/m$ associated with forming a crystalline state.)  However, in a system of interlayer excitons such high densities are generally not accessible: at much smaller densities $n \sim 1/(\max\{a_e, d\})^2$ the exciton wavefunctions overlap strongly with each other, and the individual excitons dissociate to form a state that resembles a uniform electron gas parallel to a uniform hole gas.  Thus, at $k_0 = 0$ only liquid phases are typically possible --- including, for example, a Bose-Einstein condensate at small enough $d$.\cite{butov_exciton_2003}

Consider, however, that as $k_0$ is increased from zero by the application of a perpendicular electric field, the dispersion relation of the excitons begins to flatten, as illustrated in Fig.\ \ref{fig:dispersions}.  This flat dispersion implies that the quantum confinement energy associated with forming a crystalline state is much reduced.  Correspondingly, a Wigner crystal state becomes energetically favored over a uniform state at sufficiently low density as $k_0 a_e$ approaches unity.  In other words, one can drive a liquid-to-solid transition at fixed density by increasing a transverse electric field.

The feasibility of this transition can be checked numerically in a simple way by examining a trial many-body wavefunction of dipolar bosons arranged on a triangular lattice (as employed, for example, in Ref.\ \cite{skinner_chemical_2015}).  In this approach, each lattice site is taken to be the locus of a Gaussian wavepacket of particle density, and the width $w$ of the wave packet is used as a variational parameter.  For such a wavefunction, the phase of the system can be estimated by examining the Lindemann ratio $\eta = w/\ell$, where $\ell = (2/\sqrt{3})^{1/2} n^{-1/2}$ is the lattice spacing.  Large values of $\eta$ correspond to a liquid state, while small $\eta$ suggests a Wigner crystal state.  Details of this calculation are presented in the Supplementary Information.

\begin{figure}[htb]
\centering
\includegraphics[width=0.5 \textwidth]{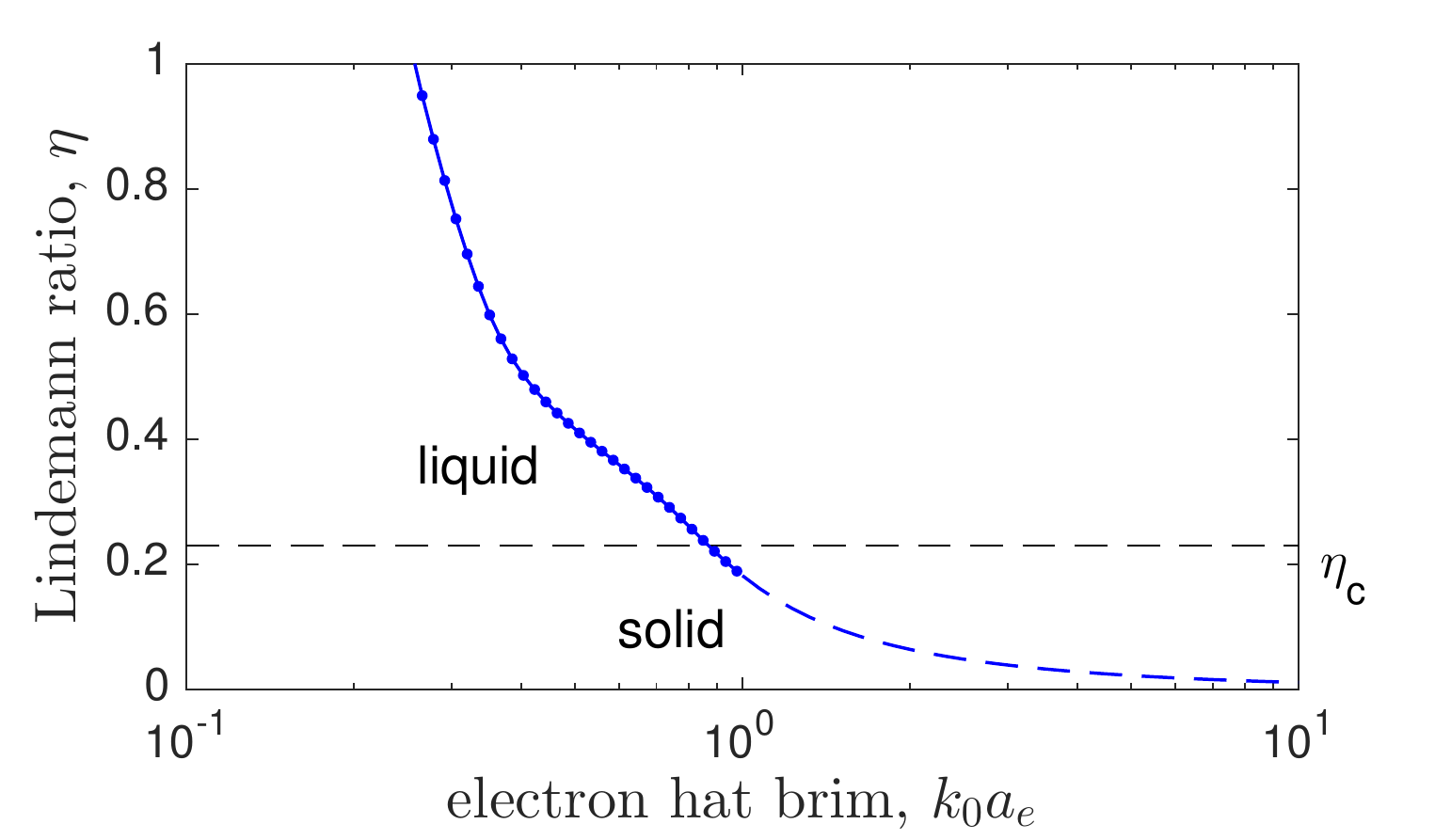}
\caption{Lindemann ratio $\eta$ for a trial wavefunction of dipolar bosons (solid line), calculated using the dispersion relation derived numerically for interlayer excitons.  Parameter values are $m_e = m_h$, $d = 0.1 a_e$, and $n = 0.01/a_e^2$. As the value of $k_0$ is increased, the system demonstrates a tendency toward Wigner crystallization (small $\eta$).  The dotted line shows the value $\eta = \eta_\text{c} \approx 0.23$ corresponding to the liquid-solid transition.\cite{babadi_universal_2013} Large $k_0 a_e \gg 1$ (dashed line) corresponds to bosons with a strongly Mexican hat-shaped dispersion, for which more exotic states may exist that are not captured by the simple variational wavefunction employed here.}
\label{fig:lindemann}
\end{figure}

Results from such a calculation are shown in Fig.\ \ref{fig:lindemann} for the example case $m_e = m_h$, $d = 0.1 a_e$, and $n = 0.01/a_e^2$.  As the value of $k_0$ is increased from zero, the Lindemann ratio declines, suggesting a tendency toward Wigner crystallization.  

If $k_0$ is increased even further, so that $k_0a_e \gg 1$, the dispersion relation of the excitons acquires a deep and wide-brimmed Mexican hat shape.  The phase of bosons with such a dispersion has been the subject of a significant amount of theoretical attention in recent years.\cite{gopalakrishnan_universal_2011, berg_electronic_2012, sedrakyan_composite_2012, sedrakyan_absence_2014, sedrakyan_statistical_2015, radic_strong_2015}  Among the proposed candidates for the ground state are a variety of exotic quantum states, including a fragmented Bose condensate,\cite{gopalakrishnan_universal_2011} a strongly anisotropic Wigner crystal,\cite{berg_electronic_2012} and a ``fermionized" state where bosons effectively exhibit Fermi statistics.\cite{sedrakyan_composite_2012, sedrakyan_absence_2014, sedrakyan_statistical_2015}  Adjudicating between these possible states is beyond the scope of the present paper, but it is interesting to note that the presence of a dipole-dipole interaction between bosons may lead to unusual transitions between the different candidate states as a function of the parameters $n$, $d$, and $k_0$.  For example, the fermionized state suggested in Refs.\ \cite{sedrakyan_absence_2014, sedrakyan_statistical_2015} for Mexican hat bosons with a delta-function interaction cannot persist at arbitrarily low density, since its chemical potential $\mu \propto n^2/m k_0^2$ is below the typical strength of the dipole-dipole interaction $\sim e^2 d^2 n^{3/2}/\kappa$.  

Nonetheless, even in the presence of a dipole interaction a number of proposed states have energy below that of a simple Bose-Einstein condensate at low density, and so there must be some kind of phase transition as a function of $k_0$.  The nature of this transition may provide fruitful ground for future studies, both theoretical and experimental.  While the majority of experimental proposals so far for realizing Mexican hat bosons have involved optically-driven Floquet bands or cold atomic gases,\cite{manchon_new_2015, sedrakyan_statistical_2015} the present work suggests that solid state bilayers or interfaces may be used as easily tunable platforms for studying the same physics, without the need for optical excitation.

\begin{figure}[htb]
\centering
\includegraphics[width=0.48 \textwidth]{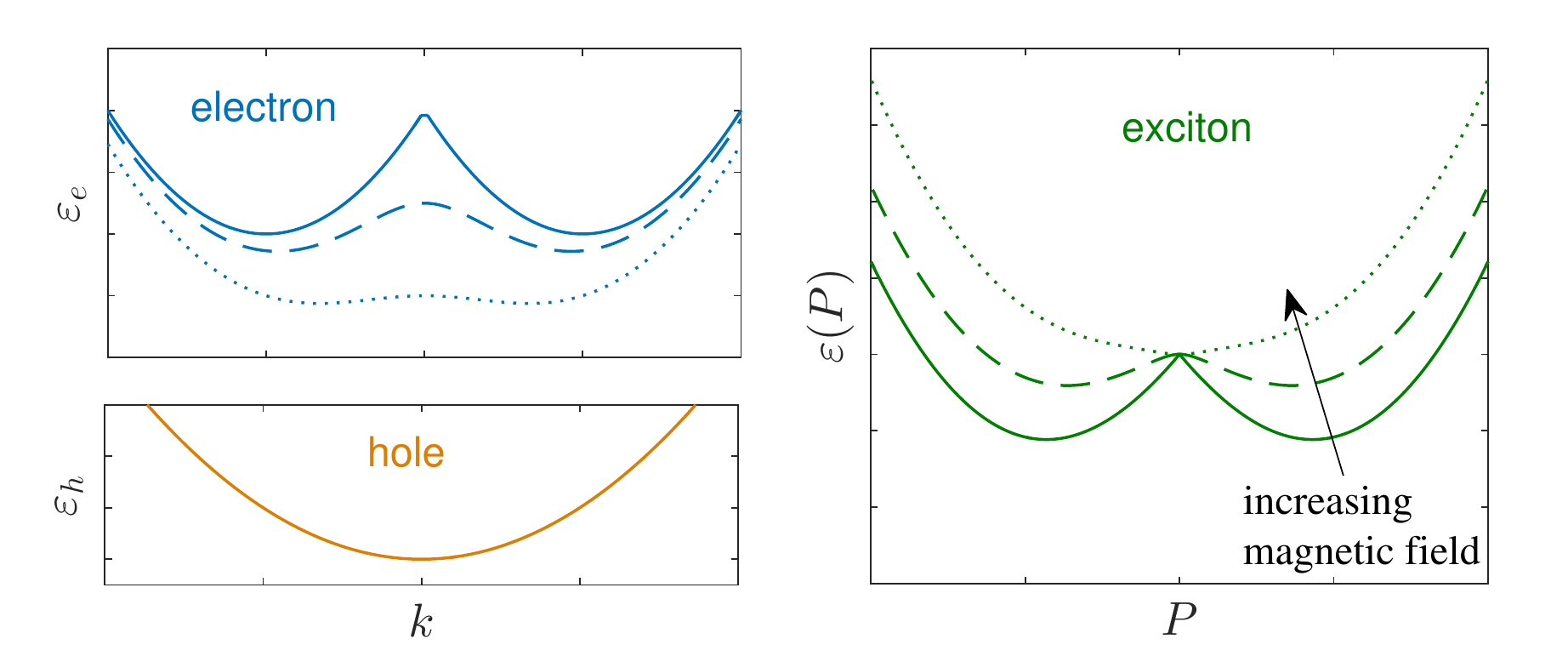}
\caption{(Color online) Schematic illustration of an alternate scheme for realizing a tunable exciton dispersion. For an electron dispersion with strong Rashba spin-orbit coupling (top left), the energy at small $k$ can be modified by application of a magnetic field.  (Here, the lower curves correspond to increasingly high magnetic field values.)  If such an electron dispersion is combined with a parabolic hole dispersion (lower left), the resulting exciton can have a dispersion relation $\e(P)$ that has a range of different forms (right).}
\label{fig:SOschematic}
\end{figure}

Finally, it is worth mentioning that in principle one can create a tunable excitonic dispersion relation even if the value of $k_0$ is fixed, provided that $k_0$ is large enough and that the small-$\kk$ part of the electron spectrum can be tuned using some other parameter.  For example, consider a system in which the electrons experience a strong Rashba spin orbit coupling.  If an external \emph{magnetic} field is applied to such a system, the spin-split kinetic energy bands hybridize and the sharp maximum in energy in the lower band is reduced, as illustrated in Fig.\ \ref{fig:SOschematic}.  If this mechanism is used to tune the electron dispersion relation, then a similar tunability can be achieved for the exciton dispersion relation.\footnote{G.\ Refael, private communication (2015).}

\acknowledgments 

\noindent
\emph{Acknowledgments.}\; I am grateful to S.\ Gopalakrishnan and G.\ Refael for valuable discussions, and to A.\ Nahum and I.\ Sodemann for critical reading of the manuscript.  This project was supported as part of the MIT Center for Excitonics, an Energy Frontier Research Center funded by the U.S. Department of Energy, Office of Science, Basic Energy Sciences under Award no. DE-SC0001088.

%

\widetext
\clearpage
\begin{center}
\textbf{\large Supplementary Information for ``Interlayer excitons with tunable dispersion relation"}
\end{center}
\setcounter{equation}{0}
\setcounter{figure}{0}
\setcounter{table}{0}
\setcounter{page}{1}
\makeatletter
\renewcommand{\theequation}{S\arabic{equation}}
\renewcommand{\thefigure}{S\arabic{figure}}
\renewcommand{\bibnumfmt}[1]{[S#1]}
\renewcommand{\citenumfont}[1]{S#1}

\begin{center}
Brian Skinner \\
\textit{Massachusetts Institute of Technology, Cambridge, MA 02139  USA} \\
(Dated: January 31, 2016)
\end{center}

\section{Kinetic and potential energy of a single exciton}
\label{sec:energy}

In Eqs.\ (10) and (11) of the main text, expressions are presented for the expectation values of the kinetic and potential energy of an exciton described by the wave function
\be 
\psi(\rr_e, \rr_h) = \exp[i (\kk_e^0 \cdot \rr_e + \kk_h^0 \cdot \rr_h)] \varphi(\rr_{eh}),
\ee
where
\be 
\varphi(x, y) = \sqrt{\frac{2 \beta^2 \lambda}{\pi}} \exp[-\beta \sqrt{x^2 + (\lambda y)^2} ].
\label{eq:psirel}
\ee 
The corresponding wave function $\pt(\kk_e, \kk_h)$ in momentum space, defined by
\be 
\psi(\rr_e, \rr_h) = \frac{1}{(2\pi)^4} \int d^2 \kk_e d^2 \kk_h \exp[i \kk_e \cdot \rr_e + i \kk_h \cdot \rr_h] \pt(\kk_e, \kk_h),
\label{eq:psir}
\ee 
is
\be 
\pt(\kk_e, \kk_h) =  \frac{16 (2 \pi)^{5/2} \beta^2}{\sqrt{\lambda}} \frac{\delta^2\left( \kk_e + \kk_h - \kk_e^0 - \kk_h^0 \right)}{\left[ 4 \beta^2 + (k_{ex} - k_{hx} - k_e^0 + k_h^0)^2 + (k_{ey} - k_{hy})^2/\lambda^2 \right]^{3/2} }.
\label{eq:psik}
\ee 
Here, $\kk_e = k_{ex} \hat{x} + k_{ey} \hat{y}$ is the electron momentum, and $\kk_h = k_{hx} \hat{x} + k_{hy} \hat{y}$ is the hole momentum.

The expectation value of the kinetic energy is given by
\be 
\e_k = \int d^2 \rr_e d^2 \rr_h \psi^*(\rr_e, \rr_h) \left( \he_e + \he_h \right) \psi(\rr_e, \rr_h),
\ee 
where $\he_e$ and $\he_h$ are the kinetic energy operators for the electron and hole, respectively.  Inserting Eq.\ (\ref{eq:psir}) into this expression, and using the fact that plane waves are eigenstates of the kinetic energy operators --- i.e., that
\begin{align}
\he_e \exp[i \kk \cdot \rr_e] & = \e_e(\kk) \exp[i \kk \cdot \rr_e], \nonumber \\
\he_h \exp[i \kk \cdot \rr_h] & = \e_h(\kk) \exp[i \kk \cdot \rr_h] \nonumber
\end{align}
--- gives the following general expression for the kinetic energy:
\be 
\e_k = \frac{1}{(2 \pi)^4} \int d^2\kk_e d^2 \kk_h \left[ \e_e(\kk_e) + \e_h(\kk_h) \right] \left| \pt(\kk_e, \kk_h) \right|^2.
\ee
Inserting Eq.\ (\ref{eq:psik}) into this expression gives Eq.\ (10) of the main text.

The potential energy of the exciton is given generally by 
\begin{align}
\e_c & = - \int d^2 \rr_e d^2 \rr_h |\psi(\rr_e, \rr_h)|^2 \frac{e^2}{\kappa \sqrt{(\rr_e - \rr_h)^2 + d^2}} \nonumber \\
& = - \int d^2 \rr_{eh} | \varphi(\rr_{eh}) |^2 \frac{e^2}{\kappa \sqrt{r_{eh}^2 + d^2}}.
\label{eq:ec}
\end{align}
Inserting Eq.\ (\ref{eq:psirel}) gives Eq.\ (11) of the main text.  In the special case $d = 0$, this integral can be done analytically, giving
\be 
e_c = - \frac{2 e^2 \beta}{\pi \kappa} \left[ K(1-1/\lambda^2) + \lambda K(1-\lambda^2) \right],
\label{eq:ecd0}
\ee
where $K(x)$ is the complete elliptic integral of the first kind. An equivalent expression to Eq.\ (\ref{eq:ecd0}) has been derived previously, for example in Refs.\ \onlinecite{schindlmayr_excitons_1997,prada_effective-mass_2015}.

\section{Energy of a trial many-body state of dipolar bosons}
\label{sec:WC}

In the main text, the feasibility of a liquid-to-solid transition driven by increasing $k_0$ was examined using a simple trial wavefunction.  Here I provide some details about this calculation, which largely follows the one presented in Ref.\ \onlinecite{skinner_chemical_2015}.

In this approach, the interlayer excitons are treated as bosonic particles with a dipolar interaction law and the dispersion relation $\e(k)$ that results from the procedure described in the main text.
The many-body wave function for the system is taken to be a product of Gaussian wave packets $\phi_{ij}(\rr)$, each centered around some point $\rr_{ij}$ on the triangular lattice.  These are given by
\be 
\phi_{ij}(\rr) = \frac{1}{\sqrt{\pi w^2}} \exp \left[ -\frac{|\rr - \rr_{ij}|^2}{2 w^2} \right],
\ee
so that the corresponding uncertainty in position for each particle is $\sqrt{\langle r^2 \rangle} = w$.  The Fourier transform of this wave packet is given by
\be 
\phit(\kk) = \sqrt{4 \pi w^2} \exp[-k^2 w^2/2],
\ee
and the corresponding kinetic energy per particle
\be 
\e_\text{kin} = \frac{1}{(2 \pi)^2} \int d^2\kk \, \e(\kk) \left| \phit(\kk) \right|^2.
\ee 
The value of $\e_\text{kin}$ is determined by numerically evaluating this integral for each value of $w$ and each instance of the dispersion relation $\e(k)$.  For the case where $\e(k) = (k-k_0)^2/2m$, the kinetic energy is given by
\be 
\e_\text{kin} = \frac{1}{2 m w^2} \left[ 1 + k_0 w (k_0 w - \sqrt{\pi}) \right]. \nonumber 
\ee 

The interaction energy (Hartree energy) in this state is given by\cite{skinner_effect_2013}
\be 
\e_\text{int} = \frac{n}{2} \sum_{q \in G} \Vt(\qq) \exp[- q^2 w^2/2] - \frac{1}{2 (2 \pi)^2} \int d^2 \kk \, \Vt(\qq) \exp[-q^2 w^2/2].
\label{eq:eI}
\ee 
Here, $n$ is the particle density, $G$ labels the set of all reciprocal lattice vectors of the triangular lattice, and
\be 
\Vt(\qq) = \frac{4 \pi e^2}{\kappa q} \left( 1 - \exp[-q d] \right) 
\ee 
is the interaction law between dipoles with a dipole arm $d$.  The second term on the right hand side of Eq.\ (\ref{eq:eI}) removes the self-interaction term from the sum in the first term.  The set of reciprocal lattice vectors are defined by
\be 
q_{ij} = 2 \pi \sqrt{\frac{2 n}{\sqrt{3}} (i^2 + i j + j^2)},
\ee 
where the indices $i$ and $j$ run over all integers.

In the variational method, the value of the positional uncertainty $\sqrt{\langle r^2 \rangle }$ is estimated to be equal to the value of $w$ which minimizes the total energy per particle, $\e_\text{kin} + \e_\text{int}$.


\begin{thebibliography}{31}%
\makeatletter
\providecommand \@ifxundefined [1]{%
 \@ifx{#1\undefined}
}%
\providecommand \@ifnum [1]{%
 \ifnum #1\expandafter \@firstoftwo
 \else \expandafter \@secondoftwo
 \fi
}%
\providecommand \@ifx [1]{%
 \ifx #1\expandafter \@firstoftwo
 \else \expandafter \@secondoftwo
 \fi
}%
\providecommand \natexlab [1]{#1}%
\providecommand \enquote  [1]{``#1''}%
\providecommand \bibnamefont  [1]{#1}%
\providecommand \bibfnamefont [1]{#1}%
\providecommand \citenamefont [1]{#1}%
\providecommand \href@noop [0]{\@secondoftwo}%
\providecommand \href [0]{\begingroup \@sanitize@url \@href}%
\providecommand \@href[1]{\@@startlink{#1}\@@href}%
\providecommand \@@href[1]{\endgroup#1\@@endlink}%
\providecommand \@sanitize@url [0]{\catcode `\\12\catcode `\$12\catcode
  `\&12\catcode `\#12\catcode `\^12\catcode `\_12\catcode `\%12\relax}%
\providecommand \@@startlink[1]{}%
\providecommand \@@endlink[0]{}%
\providecommand \url  [0]{\begingroup\@sanitize@url \@url }%
\providecommand \@url [1]{\endgroup\@href {#1}{\urlprefix }}%
\providecommand \urlprefix  [0]{URL }%
\providecommand \Eprint [0]{\href }%
\providecommand \doibase [0]{http://dx.doi.org/}%
\providecommand \selectlanguage [0]{\@gobble}%
\providecommand \bibinfo  [0]{\@secondoftwo}%
\providecommand \bibfield  [0]{\@secondoftwo}%
\providecommand \translation [1]{[#1]}%
\providecommand \BibitemOpen [0]{}%
\providecommand \bibitemStop [0]{}%
\providecommand \bibitemNoStop [0]{.\EOS\space}%
\providecommand \EOS [0]{\spacefactor3000\relax}%
\providecommand \BibitemShut  [1]{\csname bibitem#1\endcsname}%
\let\auto@bib@innerbib\@empty
\bibitem [{Note1()}]{Note1}%
  \BibitemOpen
  \bibinfo {note} {This tunability is similar in spirit to the problem of
  mobile magnetic impurities in a superfluid\cite
  {gopalakrishnan_mobile_2015}}\BibitemShut {NoStop}%
\bibitem [{\citenamefont {Lozovik}\ and\ \citenamefont
  {Yudson}(1975)}]{lozovik_feasibility_1975}%
  \BibitemOpen
  \bibfield  {author} {\bibinfo {author} {\bibfnamefont {Yu~E.}\ \bibnamefont
  {Lozovik}}\ and\ \bibinfo {author} {\bibfnamefont {V.~I.}\ \bibnamefont
  {Yudson}},\ }\bibfield  {title} {\enquote {\bibinfo {title} {Feasibility of
  superfluidity of paired spatially separated electrons and holes; a new
  superconductivity mechanism},}\ }\href
  {http://www.jetpletters.ac.ru/ps/1530/article_23400.pdf} {\bibfield
  {journal} {\bibinfo  {journal} {JETP Lett.}\ }\textbf {\bibinfo {volume}
  {22}},\ \bibinfo {pages} {274--276} (\bibinfo {year} {1975})}\BibitemShut
  {NoStop}%
\bibitem [{\citenamefont {Butov}\ and\ \citenamefont
  {Filin}(1998)}]{butov_anomalous_1998}%
  \BibitemOpen
  \bibfield  {author} {\bibinfo {author} {\bibfnamefont {L.~V.}\ \bibnamefont
  {Butov}}\ and\ \bibinfo {author} {\bibfnamefont {A.~I.}\ \bibnamefont
  {Filin}},\ }\bibfield  {title} {\enquote {\bibinfo {title} {Anomalous
  transport and luminescence of indirect excitons in {AlAs}/{GaAs} coupled
  quantum wells as evidence for exciton condensation},}\ }\href {\doibase
  10.1103/PhysRevB.58.1980} {\bibfield  {journal} {\bibinfo  {journal}
  {Physical Review B}\ }\textbf {\bibinfo {volume} {58}},\ \bibinfo {pages}
  {1980--2000} (\bibinfo {year} {1998})}\BibitemShut {NoStop}%
\bibitem [{\citenamefont {Eisenstein}\ and\ \citenamefont
  {MacDonald}(2004)}]{eisenstein_boseeinstein_2004}%
  \BibitemOpen
  \bibfield  {author} {\bibinfo {author} {\bibfnamefont {J.~P.}\ \bibnamefont
  {Eisenstein}}\ and\ \bibinfo {author} {\bibfnamefont {A.~H.}\ \bibnamefont
  {MacDonald}},\ }\bibfield  {title} {\enquote {\bibinfo {title}
  {Bose-{Einstein} condensation of excitons in bilayer electron systems},}\
  }\href {\doibase 10.1038/nature03081} {\bibfield  {journal} {\bibinfo
  {journal} {Nature}\ }\textbf {\bibinfo {volume} {432}},\ \bibinfo {pages}
  {691--694} (\bibinfo {year} {2004})}\BibitemShut {NoStop}%
\bibitem [{\citenamefont {Butov}\ \emph {et~al.}(2002)\citenamefont {Butov},
  \citenamefont {Gossard},\ and\ \citenamefont
  {Chemla}}]{butov_macroscopically_2002}%
  \BibitemOpen
  \bibfield  {author} {\bibinfo {author} {\bibfnamefont {L.~V.}\ \bibnamefont
  {Butov}}, \bibinfo {author} {\bibfnamefont {A.~C.}\ \bibnamefont {Gossard}},
  \ and\ \bibinfo {author} {\bibfnamefont {D.~S.}\ \bibnamefont {Chemla}},\
  }\bibfield  {title} {\enquote {\bibinfo {title} {Macroscopically ordered
  state in an exciton system},}\ }\href {\doibase 10.1038/nature00943}
  {\bibfield  {journal} {\bibinfo  {journal} {Nature}\ }\textbf {\bibinfo
  {volume} {418}},\ \bibinfo {pages} {751--754} (\bibinfo {year}
  {2002})}\BibitemShut {NoStop}%
\bibitem [{\citenamefont {Zhang}\ and\ \citenamefont
  {Joglekar}(2008)}]{zhang_excitonic_2008}%
  \BibitemOpen
  \bibfield  {author} {\bibinfo {author} {\bibfnamefont {C.-H.}\ \bibnamefont
  {Zhang}}\ and\ \bibinfo {author} {\bibfnamefont {Yogesh~N.}\ \bibnamefont
  {Joglekar}},\ }\bibfield  {title} {\enquote {\bibinfo {title} {Excitonic
  condensation of massless fermions in graphene bilayers},}\ }\href {\doibase
  10.1103/PhysRevB.77.233405} {\bibfield  {journal} {\bibinfo  {journal}
  {Physical Review B}\ }\textbf {\bibinfo {volume} {77}},\ \bibinfo {pages}
  {233405} (\bibinfo {year} {2008})}\BibitemShut {NoStop}%
\bibitem [{\citenamefont {Seradjeh}\ \emph {et~al.}(2009)\citenamefont
  {Seradjeh}, \citenamefont {Moore},\ and\ \citenamefont
  {Franz}}]{seradjeh_exciton_2009}%
  \BibitemOpen
  \bibfield  {author} {\bibinfo {author} {\bibfnamefont {B.}~\bibnamefont
  {Seradjeh}}, \bibinfo {author} {\bibfnamefont {J.~E.}\ \bibnamefont {Moore}},
  \ and\ \bibinfo {author} {\bibfnamefont {M.}~\bibnamefont {Franz}},\
  }\bibfield  {title} {\enquote {\bibinfo {title} {Exciton {Condensation} and
  {Charge} {Fractionalization} in a {Topological} {Insulator} {Film}},}\ }\href
  {\doibase 10.1103/PhysRevLett.103.066402} {\bibfield  {journal} {\bibinfo
  {journal} {Physical Review Letters}\ }\textbf {\bibinfo {volume} {103}},\
  \bibinfo {pages} {066402} (\bibinfo {year} {2009})}\BibitemShut {NoStop}%
\bibitem [{\citenamefont {Rivera}\ \emph {et~al.}(2015)\citenamefont {Rivera},
  \citenamefont {Schaibley}, \citenamefont {Jones}, \citenamefont {Ross},
  \citenamefont {Wu}, \citenamefont {Aivazian}, \citenamefont {Klement},
  \citenamefont {Seyler}, \citenamefont {Clark}, \citenamefont {Ghimire},
  \citenamefont {Yan}, \citenamefont {Mandrus}, \citenamefont {Yao},\ and\
  \citenamefont {Xu}}]{rivera_observation_2015}%
  \BibitemOpen
  \bibfield  {author} {\bibinfo {author} {\bibfnamefont {Pasqual}\ \bibnamefont
  {Rivera}}, \bibinfo {author} {\bibfnamefont {John~R.}\ \bibnamefont
  {Schaibley}}, \bibinfo {author} {\bibfnamefont {Aaron~M.}\ \bibnamefont
  {Jones}}, \bibinfo {author} {\bibfnamefont {Jason~S.}\ \bibnamefont {Ross}},
  \bibinfo {author} {\bibfnamefont {Sanfeng}\ \bibnamefont {Wu}}, \bibinfo
  {author} {\bibfnamefont {Grant}\ \bibnamefont {Aivazian}}, \bibinfo {author}
  {\bibfnamefont {Philip}\ \bibnamefont {Klement}}, \bibinfo {author}
  {\bibfnamefont {Kyle}\ \bibnamefont {Seyler}}, \bibinfo {author}
  {\bibfnamefont {Genevieve}\ \bibnamefont {Clark}}, \bibinfo {author}
  {\bibfnamefont {Nirmal~J.}\ \bibnamefont {Ghimire}}, \bibinfo {author}
  {\bibfnamefont {Jiaqiang}\ \bibnamefont {Yan}}, \bibinfo {author}
  {\bibfnamefont {D.~G.}\ \bibnamefont {Mandrus}}, \bibinfo {author}
  {\bibfnamefont {Wang}\ \bibnamefont {Yao}}, \ and\ \bibinfo {author}
  {\bibfnamefont {Xiaodong}\ \bibnamefont {Xu}},\ }\bibfield  {title} {\enquote
  {\bibinfo {title} {Observation of long-lived interlayer excitons in monolayer
  {MoSe}$_2$-{WSe}$_2$ heterostructures},}\ }\href {\doibase
  10.1038/ncomms7242} {\bibfield  {journal} {\bibinfo  {journal} {Nature
  Communications}\ }\textbf {\bibinfo {volume} {6}},\ \bibinfo {pages} {6242}
  (\bibinfo {year} {2015})}\BibitemShut {NoStop}%
\bibitem [{\citenamefont {Manchon}\ \emph {et~al.}(2015)\citenamefont
  {Manchon}, \citenamefont {Koo}, \citenamefont {Nitta}, \citenamefont
  {Frolov},\ and\ \citenamefont {Duine}}]{manchon_new_2015}%
  \BibitemOpen
  \bibfield  {author} {\bibinfo {author} {\bibfnamefont {A.}~\bibnamefont
  {Manchon}}, \bibinfo {author} {\bibfnamefont {H.~C.}\ \bibnamefont {Koo}},
  \bibinfo {author} {\bibfnamefont {J.}~\bibnamefont {Nitta}}, \bibinfo
  {author} {\bibfnamefont {S.~M.}\ \bibnamefont {Frolov}}, \ and\ \bibinfo
  {author} {\bibfnamefont {R.~A.}\ \bibnamefont {Duine}},\ }\bibfield  {title}
  {\enquote {\bibinfo {title} {New perspectives for {Rashba} spin-orbit
  coupling},}\ }\href {\doibase 10.1038/nmat4360} {\bibfield  {journal}
  {\bibinfo  {journal} {Nature Materials}\ }\textbf {\bibinfo {volume} {14}},\
  \bibinfo {pages} {871--882} (\bibinfo {year} {2015})}\BibitemShut {NoStop}%
\bibitem [{\citenamefont {Wang}\ \emph {et~al.}(2015)\citenamefont {Wang},
  \citenamefont {Li},\ and\ \citenamefont {Fu}}]{wang_electrical_2015}%
  \BibitemOpen
  \bibfield  {author} {\bibinfo {author} {\bibfnamefont {W.}~\bibnamefont
  {Wang}}, \bibinfo {author} {\bibfnamefont {X.~M.}\ \bibnamefont {Li}}, \ and\
  \bibinfo {author} {\bibfnamefont {J.~Y.}\ \bibnamefont {Fu}},\ }\bibfield
  {title} {\enquote {\bibinfo {title} {Electrical control of the spin-orbit
  coupling in {GaAs} from single to double and triple wells},}\ }\href
  {\doibase 10.1016/j.spmi.2015.08.021} {\bibfield  {journal} {\bibinfo
  {journal} {Superlattices and Microstructures}\ }\textbf {\bibinfo {volume}
  {88}},\ \bibinfo {pages} {43--49} (\bibinfo {year} {2015})}\BibitemShut
  {NoStop}%
\bibitem [{\citenamefont {Nitta}\ \emph {et~al.}(1997)\citenamefont {Nitta},
  \citenamefont {Akazaki}, \citenamefont {Takayanagi},\ and\ \citenamefont
  {Enoki}}]{nitta_gate_1997}%
  \BibitemOpen
  \bibfield  {author} {\bibinfo {author} {\bibfnamefont {Junsaku}\ \bibnamefont
  {Nitta}}, \bibinfo {author} {\bibfnamefont {Tatsushi}\ \bibnamefont
  {Akazaki}}, \bibinfo {author} {\bibfnamefont {Hideaki}\ \bibnamefont
  {Takayanagi}}, \ and\ \bibinfo {author} {\bibfnamefont {Takatomo}\
  \bibnamefont {Enoki}},\ }\bibfield  {title} {\enquote {\bibinfo {title} {Gate
  {Control} of {Spin}-{Orbit} {Interaction} in an {Inverted}
  {In}$_{0.53}${Ga}$_{0.47}${As}/{In}$_{0.53}$al$_{0.48}${As}
  {Heterostructure}},}\ }\href {\doibase 10.1103/PhysRevLett.78.1335}
  {\bibfield  {journal} {\bibinfo  {journal} {Physical Review Letters}\
  }\textbf {\bibinfo {volume} {78}},\ \bibinfo {pages} {1335--1338} (\bibinfo
  {year} {1997})}\BibitemShut {NoStop}%
\bibitem [{\citenamefont {Grundler}(2000)}]{grundler_large_2000}%
  \BibitemOpen
  \bibfield  {author} {\bibinfo {author} {\bibfnamefont {Dirk}\ \bibnamefont
  {Grundler}},\ }\bibfield  {title} {\enquote {\bibinfo {title} {Large {Rashba}
  {Splitting} in {InAs} {Quantum} {Wells} due to {Electron} {Wave} {Function}
  {Penetration} into the {Barrier} {Layers}},}\ }\href {\doibase
  10.1103/PhysRevLett.84.6074} {\bibfield  {journal} {\bibinfo  {journal}
  {Physical Review Letters}\ }\textbf {\bibinfo {volume} {84}},\ \bibinfo
  {pages} {6074--6077} (\bibinfo {year} {2000})}\BibitemShut {NoStop}%
\bibitem [{\citenamefont {King}\ \emph {et~al.}(2011)\citenamefont {King},
  \citenamefont {Hatch}, \citenamefont {Bianchi}, \citenamefont {Ovsyannikov},
  \citenamefont {Lupulescu}, \citenamefont {Landolt}, \citenamefont {Slomski},
  \citenamefont {Dil}, \citenamefont {Guan}, \citenamefont {Mi}, \citenamefont
  {Rienks}, \citenamefont {Fink}, \citenamefont {Lindblad}, \citenamefont
  {Svensson}, \citenamefont {Bao}, \citenamefont {Balakrishnan}, \citenamefont
  {Iversen}, \citenamefont {Osterwalder}, \citenamefont {Eberhardt},
  \citenamefont {Baumberger},\ and\ \citenamefont {Hofmann}}]{king_large_2011}%
  \BibitemOpen
  \bibfield  {author} {\bibinfo {author} {\bibfnamefont {P.~D.~C.}\
  \bibnamefont {King}}, \bibinfo {author} {\bibfnamefont {R.~C.}\ \bibnamefont
  {Hatch}}, \bibinfo {author} {\bibfnamefont {M.}~\bibnamefont {Bianchi}},
  \bibinfo {author} {\bibfnamefont {R.}~\bibnamefont {Ovsyannikov}}, \bibinfo
  {author} {\bibfnamefont {C.}~\bibnamefont {Lupulescu}}, \bibinfo {author}
  {\bibfnamefont {G.}~\bibnamefont {Landolt}}, \bibinfo {author} {\bibfnamefont
  {B.}~\bibnamefont {Slomski}}, \bibinfo {author} {\bibfnamefont {J.~H.}\
  \bibnamefont {Dil}}, \bibinfo {author} {\bibfnamefont {D.}~\bibnamefont
  {Guan}}, \bibinfo {author} {\bibfnamefont {J.~L.}\ \bibnamefont {Mi}},
  \bibinfo {author} {\bibfnamefont {E.~D.~L.}\ \bibnamefont {Rienks}}, \bibinfo
  {author} {\bibfnamefont {J.}~\bibnamefont {Fink}}, \bibinfo {author}
  {\bibfnamefont {A.}~\bibnamefont {Lindblad}}, \bibinfo {author}
  {\bibfnamefont {S.}~\bibnamefont {Svensson}}, \bibinfo {author}
  {\bibfnamefont {S.}~\bibnamefont {Bao}}, \bibinfo {author} {\bibfnamefont
  {G.}~\bibnamefont {Balakrishnan}}, \bibinfo {author} {\bibfnamefont {B.~B.}\
  \bibnamefont {Iversen}}, \bibinfo {author} {\bibfnamefont {J.}~\bibnamefont
  {Osterwalder}}, \bibinfo {author} {\bibfnamefont {W.}~\bibnamefont
  {Eberhardt}}, \bibinfo {author} {\bibfnamefont {F.}~\bibnamefont
  {Baumberger}}, \ and\ \bibinfo {author} {\bibfnamefont {Ph.}\ \bibnamefont
  {Hofmann}},\ }\bibfield  {title} {\enquote {\bibinfo {title} {Large {Tunable}
  {Rashba} {Spin} {Splitting} of a {Two}-{Dimensional} {Electron} {Gas} in
  {Bi}$_2${Se}$_3$},}\ }\href {\doibase 10.1103/PhysRevLett.107.096802}
  {\bibfield  {journal} {\bibinfo  {journal} {Physical Review Letters}\
  }\textbf {\bibinfo {volume} {107}},\ \bibinfo {pages} {096802} (\bibinfo
  {year} {2011})}\BibitemShut {NoStop}%
\bibitem [{\citenamefont {Caviglia}\ \emph {et~al.}(2010)\citenamefont
  {Caviglia}, \citenamefont {Gabay}, \citenamefont {Gariglio}, \citenamefont
  {Reyren}, \citenamefont {Cancellieri},\ and\ \citenamefont
  {Triscone}}]{caviglia_tunable_2010}%
  \BibitemOpen
  \bibfield  {author} {\bibinfo {author} {\bibfnamefont {A.~D.}\ \bibnamefont
  {Caviglia}}, \bibinfo {author} {\bibfnamefont {M.}~\bibnamefont {Gabay}},
  \bibinfo {author} {\bibfnamefont {S.}~\bibnamefont {Gariglio}}, \bibinfo
  {author} {\bibfnamefont {N.}~\bibnamefont {Reyren}}, \bibinfo {author}
  {\bibfnamefont {C.}~\bibnamefont {Cancellieri}}, \ and\ \bibinfo {author}
  {\bibfnamefont {J.-M.}\ \bibnamefont {Triscone}},\ }\bibfield  {title}
  {\enquote {\bibinfo {title} {Tunable {Rashba} {Spin}-{Orbit} {Interaction} at
  {Oxide} {Interfaces}},}\ }\href {\doibase 10.1103/PhysRevLett.104.126803}
  {\bibfield  {journal} {\bibinfo  {journal} {Physical Review Letters}\
  }\textbf {\bibinfo {volume} {104}},\ \bibinfo {pages} {126803} (\bibinfo
  {year} {2010})}\BibitemShut {NoStop}%
\bibitem [{\citenamefont {McCann}\ and\ \citenamefont
  {Fal'ko}(2006)}]{mccann_landau-level_2006}%
  \BibitemOpen
  \bibfield  {author} {\bibinfo {author} {\bibfnamefont {Edward}\ \bibnamefont
  {McCann}}\ and\ \bibinfo {author} {\bibfnamefont {Vladimir~I.}\ \bibnamefont
  {Fal'ko}},\ }\bibfield  {title} {\enquote {\bibinfo {title} {Landau-{Level}
  {Degeneracy} and {Quantum} {Hall} {Effect} in a {Graphite} {Bilayer}},}\
  }\href {\doibase 10.1103/PhysRevLett.96.086805} {\bibfield  {journal}
  {\bibinfo  {journal} {Phys. Rev. Lett.}\ }\textbf {\bibinfo {volume} {96}},\
  \bibinfo {pages} {086805} (\bibinfo {year} {2006})}\BibitemShut {NoStop}%
\bibitem [{\citenamefont {McCann}\ and\ \citenamefont
  {Koshino}(2013)}]{mccann_electronic_2013}%
  \BibitemOpen
  \bibfield  {author} {\bibinfo {author} {\bibfnamefont {Edward}\ \bibnamefont
  {McCann}}\ and\ \bibinfo {author} {\bibfnamefont {Mikito}\ \bibnamefont
  {Koshino}},\ }\bibfield  {title} {\enquote {\bibinfo {title} {The electronic
  properties of bilayer graphene},}\ }\href
  {http://stacks.iop.org/0034-4885/76/i=5/a=056503} {\bibfield  {journal}
  {\bibinfo  {journal} {Reports on Progress in Physics}\ }\textbf {\bibinfo
  {volume} {76}},\ \bibinfo {pages} {056503} (\bibinfo {year}
  {2013})}\BibitemShut {NoStop}%
\bibitem [{\citenamefont {Skinner}\ \emph {et~al.}(2014)\citenamefont
  {Skinner}, \citenamefont {Shklovskii},\ and\ \citenamefont
  {Voloshin}}]{skinner_bound_2014}%
  \BibitemOpen
  \bibfield  {author} {\bibinfo {author} {\bibfnamefont {Brian}\ \bibnamefont
  {Skinner}}, \bibinfo {author} {\bibfnamefont {B.~I.}\ \bibnamefont
  {Shklovskii}}, \ and\ \bibinfo {author} {\bibfnamefont {M.~B.}\ \bibnamefont
  {Voloshin}},\ }\bibfield  {title} {\enquote {\bibinfo {title} {Bound state
  energy of a {Coulomb} impurity in gapped bilayer graphene},}\ }\href
  {\doibase 10.1103/PhysRevB.89.041405} {\bibfield  {journal} {\bibinfo
  {journal} {Physical Review B}\ }\textbf {\bibinfo {volume} {89}},\ \bibinfo
  {pages} {041405} (\bibinfo {year} {2014})}\BibitemShut {NoStop}%
\bibitem [{\citenamefont {Kittel}\ and\ \citenamefont
  {Mitchell}(1954)}]{kittel_theory_1954}%
  \BibitemOpen
  \bibfield  {author} {\bibinfo {author} {\bibfnamefont {C.}~\bibnamefont
  {Kittel}}\ and\ \bibinfo {author} {\bibfnamefont {A.~H.}\ \bibnamefont
  {Mitchell}},\ }\bibfield  {title} {\enquote {\bibinfo {title} {Theory of
  {Donor} and {Acceptor} {States} in {Silicon} and {Germanium}},}\ }\href
  {\doibase 10.1103/PhysRev.96.1488} {\bibfield  {journal} {\bibinfo  {journal}
  {Physical Review}\ }\textbf {\bibinfo {volume} {96}},\ \bibinfo {pages}
  {1488--1493} (\bibinfo {year} {1954})}\BibitemShut {NoStop}%
\bibitem [{\citenamefont {Prada}\ \emph {et~al.}(2015)\citenamefont {Prada},
  \citenamefont {Alvarez}, \citenamefont {Narasimha-Acharya}, \citenamefont
  {Bailen},\ and\ \citenamefont {Palacios}}]{prada_effective-mass_2015}%
  \BibitemOpen
  \bibfield  {author} {\bibinfo {author} {\bibfnamefont {Elsa}\ \bibnamefont
  {Prada}}, \bibinfo {author} {\bibfnamefont {J.~V.}\ \bibnamefont {Alvarez}},
  \bibinfo {author} {\bibfnamefont {K.~L.}\ \bibnamefont {Narasimha-Acharya}},
  \bibinfo {author} {\bibfnamefont {F.~J.}\ \bibnamefont {Bailen}}, \ and\
  \bibinfo {author} {\bibfnamefont {J.~J.}\ \bibnamefont {Palacios}},\
  }\bibfield  {title} {\enquote {\bibinfo {title} {Effective-mass theory for
  the anisotropic exciton in two-dimensional crystals: {Application} to
  phosphorene},}\ }\href {\doibase 10.1103/PhysRevB.91.245421} {\bibfield
  {journal} {\bibinfo  {journal} {Physical Review B}\ }\textbf {\bibinfo
  {volume} {91}},\ \bibinfo {pages} {245421} (\bibinfo {year}
  {2015})}\BibitemShut {NoStop}%
\bibitem [{\citenamefont {Babadi}\ \emph {et~al.}(2013)\citenamefont {Babadi},
  \citenamefont {Skinner}, \citenamefont {Fogler},\ and\ \citenamefont
  {Demler}}]{babadi_universal_2013}%
  \BibitemOpen
  \bibfield  {author} {\bibinfo {author} {\bibfnamefont {Mehrtash}\
  \bibnamefont {Babadi}}, \bibinfo {author} {\bibfnamefont {Brian}\
  \bibnamefont {Skinner}}, \bibinfo {author} {\bibfnamefont {Michael~M.}\
  \bibnamefont {Fogler}}, \ and\ \bibinfo {author} {\bibfnamefont {Eugene}\
  \bibnamefont {Demler}},\ }\bibfield  {title} {\enquote {\bibinfo {title}
  {Universal behavior of repulsive two-dimensional fermions in the vicinity of
  the quantum freezing point},}\ }\href {\doibase 10.1209/0295-5075/103/16002}
  {\bibfield  {journal} {\bibinfo  {journal} {EPL (Europhysics Letters)}\
  }\textbf {\bibinfo {volume} {103}},\ \bibinfo {pages} {16002} (\bibinfo
  {year} {2013})}\BibitemShut {NoStop}%
\bibitem [{\citenamefont {Astrakharchik}\ \emph {et~al.}(2007)\citenamefont
  {Astrakharchik}, \citenamefont {Boronat}, \citenamefont {Kurbakov},\ and\
  \citenamefont {Lozovik}}]{astrakharchik_quantum_2007}%
  \BibitemOpen
  \bibfield  {author} {\bibinfo {author} {\bibfnamefont {G.~E.}\ \bibnamefont
  {Astrakharchik}}, \bibinfo {author} {\bibfnamefont {J.}~\bibnamefont
  {Boronat}}, \bibinfo {author} {\bibfnamefont {I.~L.}\ \bibnamefont
  {Kurbakov}}, \ and\ \bibinfo {author} {\bibfnamefont {Yu.~E.}\ \bibnamefont
  {Lozovik}},\ }\bibfield  {title} {\enquote {\bibinfo {title} {Quantum {Phase}
  {Transition} in a {Two}-{Dimensional} {System} of {Dipoles}},}\ }\href
  {\doibase 10.1103/PhysRevLett.98.060405} {\bibfield  {journal} {\bibinfo
  {journal} {Physical Review Letters}\ }\textbf {\bibinfo {volume} {98}},\
  \bibinfo {pages} {060405} (\bibinfo {year} {2007})}\BibitemShut {NoStop}%
\bibitem [{\citenamefont {Butov}(2003)}]{butov_exciton_2003}%
  \BibitemOpen
  \bibfield  {author} {\bibinfo {author} {\bibfnamefont {L.~V.}\ \bibnamefont
  {Butov}},\ }\bibfield  {title} {\enquote {\bibinfo {title} {Exciton
  condensation in coupled quantum wells},}\ }\href {\doibase
  10.1016/S0038-1098(03)00312-0} {\bibfield  {journal} {\bibinfo  {journal}
  {Solid State Communications}\ }\bibinfo {series} {Quantum {Phases} at the
  {Nanoscale}},\ \textbf {\bibinfo {volume} {127}},\ \bibinfo {pages} {89--98}
  (\bibinfo {year} {2003})}\BibitemShut {NoStop}%
\bibitem [{\citenamefont {Skinner}(2015)}]{skinner_chemical_2015}%
  \BibitemOpen
  \bibfield  {author} {\bibinfo {author} {\bibfnamefont {Brian}\ \bibnamefont
  {Skinner}},\ }\bibfield  {title} {\enquote {\bibinfo {title} {Chemical
  potential and compressibility of quantum {Hall} bilayer excitons},}\ }\href
  {http://arxiv.org/abs/1511.03287} {\bibfield  {journal} {\bibinfo  {journal}
  {arXiv:1511.03287 [cond-mat]}\ } (\bibinfo {year} {2015})}\BibitemShut
  {NoStop}%
\bibitem [{\citenamefont {Gopalakrishnan}\ \emph {et~al.}(2011)\citenamefont
  {Gopalakrishnan}, \citenamefont {Lamacraft},\ and\ \citenamefont
  {Goldbart}}]{gopalakrishnan_universal_2011}%
  \BibitemOpen
  \bibfield  {author} {\bibinfo {author} {\bibfnamefont {Sarang}\ \bibnamefont
  {Gopalakrishnan}}, \bibinfo {author} {\bibfnamefont {Austen}\ \bibnamefont
  {Lamacraft}}, \ and\ \bibinfo {author} {\bibfnamefont {Paul~M.}\ \bibnamefont
  {Goldbart}},\ }\bibfield  {title} {\enquote {\bibinfo {title} {Universal
  phase structure of dilute {Bose} gases with {Rashba} spin-orbit coupling},}\
  }\href {\doibase 10.1103/PhysRevA.84.061604} {\bibfield  {journal} {\bibinfo
  {journal} {Physical Review A}\ }\textbf {\bibinfo {volume} {84}},\ \bibinfo
  {pages} {061604} (\bibinfo {year} {2011})}\BibitemShut {NoStop}%
\bibitem [{\citenamefont {Berg}\ \emph {et~al.}(2012)\citenamefont {Berg},
  \citenamefont {Rudner},\ and\ \citenamefont
  {Kivelson}}]{berg_electronic_2012}%
  \BibitemOpen
  \bibfield  {author} {\bibinfo {author} {\bibfnamefont {Erez}\ \bibnamefont
  {Berg}}, \bibinfo {author} {\bibfnamefont {Mark~S.}\ \bibnamefont {Rudner}},
  \ and\ \bibinfo {author} {\bibfnamefont {Steven~A.}\ \bibnamefont
  {Kivelson}},\ }\bibfield  {title} {\enquote {\bibinfo {title} {Electronic
  liquid crystalline phases in a spin-orbit coupled two-dimensional electron
  gas},}\ }\href {\doibase 10.1103/PhysRevB.85.035116} {\bibfield  {journal}
  {\bibinfo  {journal} {Physical Review B}\ }\textbf {\bibinfo {volume} {85}},\
  \bibinfo {pages} {035116} (\bibinfo {year} {2012})}\BibitemShut {NoStop}%
\bibitem [{\citenamefont {Sedrakyan}\ \emph {et~al.}(2012)\citenamefont
  {Sedrakyan}, \citenamefont {Kamenev},\ and\ \citenamefont
  {Glazman}}]{sedrakyan_composite_2012}%
  \BibitemOpen
  \bibfield  {author} {\bibinfo {author} {\bibfnamefont {Tigran~A.}\
  \bibnamefont {Sedrakyan}}, \bibinfo {author} {\bibfnamefont {Alex}\
  \bibnamefont {Kamenev}}, \ and\ \bibinfo {author} {\bibfnamefont {Leonid~I.}\
  \bibnamefont {Glazman}},\ }\bibfield  {title} {\enquote {\bibinfo {title}
  {Composite fermion state of spin-orbit-coupled bosons},}\ }\href {\doibase
  10.1103/PhysRevA.86.063639} {\bibfield  {journal} {\bibinfo  {journal}
  {Physical Review A}\ }\textbf {\bibinfo {volume} {86}},\ \bibinfo {pages}
  {063639} (\bibinfo {year} {2012})}\BibitemShut {NoStop}%
\bibitem [{\citenamefont {Sedrakyan}\ \emph {et~al.}(2014)\citenamefont
  {Sedrakyan}, \citenamefont {Glazman},\ and\ \citenamefont
  {Kamenev}}]{sedrakyan_absence_2014}%
  \BibitemOpen
  \bibfield  {author} {\bibinfo {author} {\bibfnamefont {Tigran~A.}\
  \bibnamefont {Sedrakyan}}, \bibinfo {author} {\bibfnamefont {Leonid~I.}\
  \bibnamefont {Glazman}}, \ and\ \bibinfo {author} {\bibfnamefont {Alex}\
  \bibnamefont {Kamenev}},\ }\bibfield  {title} {\enquote {\bibinfo {title}
  {Absence of {Bose} condensation on lattices with moat bands},}\ }\href
  {\doibase 10.1103/PhysRevB.89.201112} {\bibfield  {journal} {\bibinfo
  {journal} {Physical Review B}\ }\textbf {\bibinfo {volume} {89}},\ \bibinfo
  {pages} {201112} (\bibinfo {year} {2014})}\BibitemShut {NoStop}%
\bibitem [{\citenamefont {Sedrakyan}\ \emph {et~al.}(2015)\citenamefont
  {Sedrakyan}, \citenamefont {Galitski},\ and\ \citenamefont
  {Kamenev}}]{sedrakyan_statistical_2015}%
  \BibitemOpen
  \bibfield  {author} {\bibinfo {author} {\bibfnamefont {Tigran~A.}\
  \bibnamefont {Sedrakyan}}, \bibinfo {author} {\bibfnamefont {Victor~M.}\
  \bibnamefont {Galitski}}, \ and\ \bibinfo {author} {\bibfnamefont {Alex}\
  \bibnamefont {Kamenev}},\ }\bibfield  {title} {\enquote {\bibinfo {title}
  {Statistical {Transmutation} in {Floquet} {Driven} {Optical} {Lattices}},}\
  }\href {\doibase 10.1103/PhysRevLett.115.195301} {\bibfield  {journal}
  {\bibinfo  {journal} {Physical Review Letters}\ }\textbf {\bibinfo {volume}
  {115}},\ \bibinfo {pages} {195301} (\bibinfo {year} {2015})}\BibitemShut
  {NoStop}%
\bibitem [{\citenamefont {Radi\ifmmode~\acute{c}\else \'{c}\fi{}}\ \emph
  {et~al.}(2015)\citenamefont {Radi\ifmmode~\acute{c}\else \'{c}\fi{}},
  \citenamefont {Natu},\ and\ \citenamefont {Galitski}}]{radic_strong_2015}%
  \BibitemOpen
  \bibfield  {author} {\bibinfo {author} {\bibfnamefont {Juraj}\ \bibnamefont
  {Radi\ifmmode~\acute{c}\else \'{c}\fi{}}}, \bibinfo {author} {\bibfnamefont
  {Stefan~S.}\ \bibnamefont {Natu}}, \ and\ \bibinfo {author} {\bibfnamefont
  {Victor}\ \bibnamefont {Galitski}},\ }\bibfield  {title} {\enquote {\bibinfo
  {title} {Strong correlation effects in a two-dimensional bose gas with
  quartic dispersion},}\ }\href {\doibase 10.1103/PhysRevA.91.063634}
  {\bibfield  {journal} {\bibinfo  {journal} {Phys. Rev. A}\ }\textbf {\bibinfo
  {volume} {91}},\ \bibinfo {pages} {063634} (\bibinfo {year}
  {2015})}\BibitemShut {NoStop}%
\bibitem [{Note2()}]{Note2}%
  \BibitemOpen
  \bibinfo {note} {G.\ Refael, private communication (2015).}\BibitemShut
  {Stop}%
\bibitem [{\citenamefont {Gopalakrishnan}\ \emph {et~al.}(2015)\citenamefont
  {Gopalakrishnan}, \citenamefont {Parker},\ and\ \citenamefont
  {Demler}}]{gopalakrishnan_mobile_2015}%
  \BibitemOpen
  \bibfield  {author} {\bibinfo {author} {\bibfnamefont {Sarang}\ \bibnamefont
  {Gopalakrishnan}}, \bibinfo {author} {\bibfnamefont {Colin~V.}\ \bibnamefont
  {Parker}}, \ and\ \bibinfo {author} {\bibfnamefont {Eugene}\ \bibnamefont
  {Demler}},\ }\bibfield  {title} {\enquote {\bibinfo {title} {Mobile magnetic
  impurities in a fermi superfluid: A route to designer molecules},}\ }\href
  {\doibase 10.1103/PhysRevLett.114.045301} {\bibfield  {journal} {\bibinfo
  {journal} {Phys. Rev. Lett.}\ }\textbf {\bibinfo {volume} {114}},\ \bibinfo
  {pages} {045301} (\bibinfo {year} {2015})}\BibitemShut {NoStop}%
\end{thebibliography}

\begin{thebibliography}{4}%
\makeatletter
\providecommand \@ifxundefined [1]{%
 \@ifx{#1\undefined}
}%
\providecommand \@ifnum [1]{%
 \ifnum #1\expandafter \@firstoftwo
 \else \expandafter \@secondoftwo
 \fi
}%
\providecommand \@ifx [1]{%
 \ifx #1\expandafter \@firstoftwo
 \else \expandafter \@secondoftwo
 \fi
}%
\providecommand \natexlab [1]{#1}%
\providecommand \enquote  [1]{``#1''}%
\providecommand \bibnamefont  [1]{#1}%
\providecommand \bibfnamefont [1]{#1}%
\providecommand \citenamefont [1]{#1}%
\providecommand \href@noop [0]{\@secondoftwo}%
\providecommand \href [0]{\begingroup \@sanitize@url \@href}%
\providecommand \@href[1]{\@@startlink{#1}\@@href}%
\providecommand \@@href[1]{\endgroup#1\@@endlink}%
\providecommand \@sanitize@url [0]{\catcode `\\12\catcode `\$12\catcode
  `\&12\catcode `\#12\catcode `\^12\catcode `\_12\catcode `\%12\relax}%
\providecommand \@@startlink[1]{}%
\providecommand \@@endlink[0]{}%
\providecommand \url  [0]{\begingroup\@sanitize@url \@url }%
\providecommand \@url [1]{\endgroup\@href {#1}{\urlprefix }}%
\providecommand \urlprefix  [0]{URL }%
\providecommand \Eprint [0]{\href }%
\providecommand \doibase [0]{http://dx.doi.org/}%
\providecommand \selectlanguage [0]{\@gobble}%
\providecommand \bibinfo  [0]{\@secondoftwo}%
\providecommand \bibfield  [0]{\@secondoftwo}%
\providecommand \translation [1]{[#1]}%
\providecommand \BibitemOpen [0]{}%
\providecommand \bibitemStop [0]{}%
\providecommand \bibitemNoStop [0]{.\EOS\space}%
\providecommand \EOS [0]{\spacefactor3000\relax}%
\providecommand \BibitemShut  [1]{\csname bibitem#1\endcsname}%
\let\auto@bib@innerbib\@empty
\bibitem [{\citenamefont {Schindlmayr}(1997)}]{schindlmayr_excitons_1997}%
  \BibitemOpen
  \bibfield  {author} {\bibinfo {author} {\bibfnamefont {Arno}\ \bibnamefont
  {Schindlmayr}},\ }\bibfield  {title} {\enquote {\bibinfo {title} {Excitons
  with anisotropic effective mass},}\ }\href {\doibase
  10.1088/0143-0807/18/5/011} {\bibfield  {journal} {\bibinfo  {journal}
  {European Journal of Physics}\ }\textbf {\bibinfo {volume} {18}},\ \bibinfo
  {pages} {374} (\bibinfo {year} {1997})}\BibitemShut {NoStop}%
\bibitem [{\citenamefont {Prada}\ \emph {et~al.}(2015)\citenamefont {Prada},
  \citenamefont {Alvarez}, \citenamefont {Narasimha-Acharya}, \citenamefont
  {Bailen},\ and\ \citenamefont {Palacios}}]{prada_effective-mass_2015}%
  \BibitemOpen
  \bibfield  {author} {\bibinfo {author} {\bibfnamefont {Elsa}\ \bibnamefont
  {Prada}}, \bibinfo {author} {\bibfnamefont {J.~V.}\ \bibnamefont {Alvarez}},
  \bibinfo {author} {\bibfnamefont {K.~L.}\ \bibnamefont {Narasimha-Acharya}},
  \bibinfo {author} {\bibfnamefont {F.~J.}\ \bibnamefont {Bailen}}, \ and\
  \bibinfo {author} {\bibfnamefont {J.~J.}\ \bibnamefont {Palacios}},\
  }\bibfield  {title} {\enquote {\bibinfo {title} {Effective-mass theory for
  the anisotropic exciton in two-dimensional crystals: {Application} to
  phosphorene},}\ }\href {\doibase 10.1103/PhysRevB.91.245421} {\bibfield
  {journal} {\bibinfo  {journal} {Physical Review B}\ }\textbf {\bibinfo
  {volume} {91}},\ \bibinfo {pages} {245421} (\bibinfo {year}
  {2015})}\BibitemShut {NoStop}%
\bibitem [{\citenamefont {Skinner}(2015)}]{skinner_chemical_2015}%
  \BibitemOpen
  \bibfield  {author} {\bibinfo {author} {\bibfnamefont {Brian}\ \bibnamefont
  {Skinner}},\ }\bibfield  {title} {\enquote {\bibinfo {title} {Chemical
  potential and compressibility of quantum {Hall} bilayer excitons},}\ }\href
  {http://arxiv.org/abs/1511.03287} {\bibfield  {journal} {\bibinfo  {journal}
  {arXiv:1511.03287 [cond-mat]}\ } (\bibinfo {year} {2015})},\ \bibinfo {note}
  {arXiv: 1511.03287}\BibitemShut {NoStop}%
\bibitem [{\citenamefont {Skinner}\ \emph {et~al.}(2013)\citenamefont
  {Skinner}, \citenamefont {Yu}, \citenamefont {Kretinin}, \citenamefont
  {Geim}, \citenamefont {Novoselov},\ and\ \citenamefont
  {Shklovskii}}]{skinner_effect_2013}%
  \BibitemOpen
  \bibfield  {author} {\bibinfo {author} {\bibfnamefont {Brian}\ \bibnamefont
  {Skinner}}, \bibinfo {author} {\bibfnamefont {G.~L.}\ \bibnamefont {Yu}},
  \bibinfo {author} {\bibfnamefont {A.~V.}\ \bibnamefont {Kretinin}}, \bibinfo
  {author} {\bibfnamefont {A.~K.}\ \bibnamefont {Geim}}, \bibinfo {author}
  {\bibfnamefont {K.~S.}\ \bibnamefont {Novoselov}}, \ and\ \bibinfo {author}
  {\bibfnamefont {B.~I.}\ \bibnamefont {Shklovskii}},\ }\bibfield  {title}
  {\enquote {\bibinfo {title} {Effect of dielectric response on the quantum
  capacitance of graphene in a strong magnetic field},}\ }\href {\doibase
  10.1103/PhysRevB.88.155417} {\bibfield  {journal} {\bibinfo  {journal}
  {Physical Review B}\ }\textbf {\bibinfo {volume} {88}},\ \bibinfo {pages}
  {155417} (\bibinfo {year} {2013})}\BibitemShut {NoStop}%
\end{thebibliography}
\end{document}